# Thermal Study of the X+YCl (X, Y=H, D) Reactions

Li Yao,[a,b] Ke-Li Han,[b*] He-Shan Song,[a] and Dong-Hui Zhang[c]

[a]*Department of Physics, Dalian University of Technology, Dalian 116023, China*

[b]*Center for Computational Chemistry and State Key Laboratory of Molecular Reaction Dynamics, Dalian Institute of Chemical Physics, Chinese Academy of Sciences, Dalian 116023, China*

[c]*Department of Computational Science, National University of Singapore. Singapore*

**Abstract**

Time-dependent wave packet calculations for the reaction H+HCl and its isotopic reactions are carried out on the potential energy surface (PES) of Bian and Werner (BW2) [Bian, W.; Werner, H. –J., J. Chem. Phys. 2000, 112, 220]. Reaction probabilities for the exchanged and abstraction channels are calculated from various initial rotational states of the reagent. Those have then been used to estimate reaction cross sections and rate constants which also are calculated and explained by the zero-point energy and the tunneling effect. The results of this work were compared with that of previous quasiclassical trajectory calculations and reaction dynamics experiments on the abstraction channel. In addition, the calculated rate constants are in reasonably good agreement with experimental measurements for both channels.

**Key words:** *ab initio*, H/D+ DCl/HCl, collision energy, integral cross section, rate constant.

[*] Corresponding author: E-mail: klhan@ms.dicp.ac.cn

## 1. Introduction

The time-dependent quantum wave packet (TDWP) approach[1,2] has emerged as a powerful computational tool for studying quantum reaction dynamics of A+BC→AB+C systems.[2-17] Furthermore, the TDWP approach is conceptually simple. It provides an interpretation of the numerical results through time propagation of the wave packet.[7-9]

The gas-phase reaction of H+HCl and the corresponding isotopic reaction of H+DCl represent important elementary steps in the $H_2/D_2+Cl_2$→2H/DCl reaction system, which has played a major role in the development of chemical kinetics and the environment in atmospheric chemistry.[21-25] Whereas the measured relative excitations functions $\sigma_R$ ($E_{col}$), obtained in the collision energy range $E_{col}$=0.17-0.35eV, revealed a marked preference for H+DCl product channel. Further QCT calculations have been carried out for the H+DCl isotope reaction which can be compared with the results of previous reaction dynamics experiments of Barclay et al.[26] Polanyi and co-workers observed that for the H+DCl reaction the abstraction cross section decreases with increasing collision energy from a value of 0.2Å$^2$ at $E_{col}$=1.2eV to a value of 0.1Å$^2$ at $E_{col}$=1.8eV.[26]

The first globally realistic Cl-H-H potential energy surface (PES) was calculated by Baer and Last[27] and more recent PES have been published by Truhlar et. al.[28] The latter one was used by Polanyi et. al. in quasiclassical trajectory (QCT) calculations to study the branching between D atom abstraction and hydrogen atom exchanged in the reaction of translationally energetic H atoms with DCl.[26] The D atom abstraction and hydrogen atom exchange reaction cross section measurements are carried out for the H+DCl reaction in the energy range of 1.0~2.4eV.[26] In 1999 a new, completely *ab initio*, PES of Bian and Werner[29] (BW2) became available. The PES of Bian and Werner was developed based on extensive *ab initio* calculations using the highly accurate electronic structure methods and very large basis sets presently available. The *ab initio* calculations were carried out at more than 1200 nuclear geometries. It was further improved by scaling the correlation energies at all geometries with a constant factor.[29]

In this paper, we have carried out TDWP calculations for the following four different reactions,

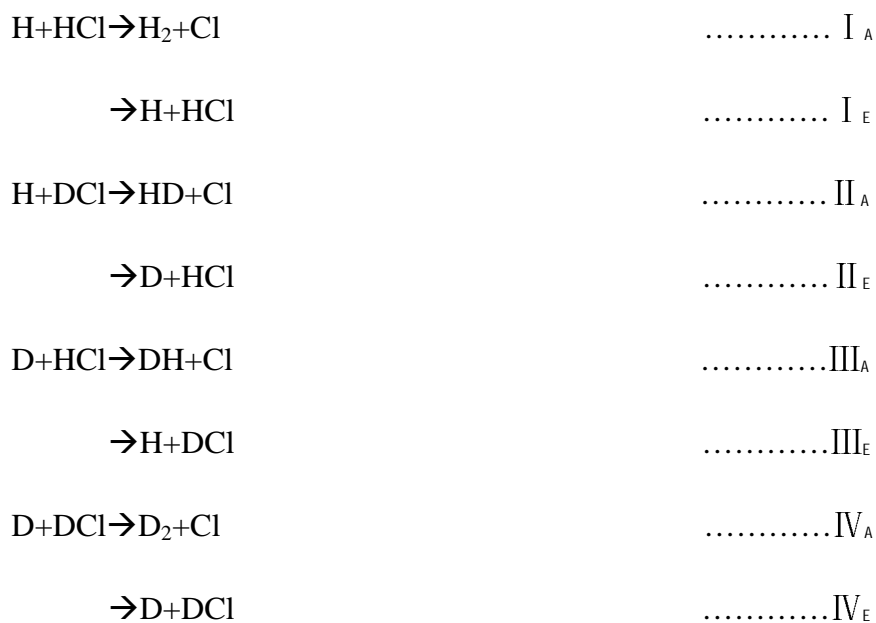

| H+HCl→H$_2$+Cl | ………… I$_A$ |
| →H+HCl | ………… I$_E$ |
| H+DCl→HD+Cl | ………… II$_A$ |
| →D+HCl | ………… II$_E$ |
| D+HCl→DH+Cl | …………III$_A$ |
| →H+DCl | …………III$_E$ |
| D+DCl→D$_2$+Cl | …………IV$_A$ |
| →D+DCl | …………IV$_E$ |

to find out the isotopic effect. Where the X$_E$ denoted the exchanged channel, and the X$_A$ denoted the abstraction channel where X denoted the Ⅰ, Ⅱ, Ⅲ, Ⅳ.

The isotopic effect in chemical reactions has attracted great interest for a long time. Quantum tunneling and zero-point energy shift are the most common isotopic effects that are usually investigated. First, when a chemical reaction occurs at low temperature and translational energy, the isotopic tunneling effect can often make the reaction happen below the classical barrier height. And this effect often appears when the effective mass of the coordinate is strongly coupled with the reaction pathway changes, e.g., generally the proton abstraction exhibits a larger tunneling effect than that of the proton replaced by deuterium. Second, for a chemical reaction process, the isotopic effect on the potential energy surface reflects mainly on the zero-point energy revision to the reaction path and this revision is caused by the change of the normal-vibration frequencies. This kind of shift sometimes can give a remarkable influence on the micro-dynamics.

In addition, the H+HCl and its isotopic variants have played an important role in chemical kinetics due to its importance to the environment in atmospheric chemistry.[25] The dynamics of the four reactions have been the subject of many theoretical and experimental studies. To the best of our knowledge, no TDWP for the H+HCl and its isotopic reactions system has been reported so far on the BW2 PES.

The aim of the present study was to investigate the collision energy dependence of the probabilities (for total angular momentum quantum number $J$), reaction cross sections and thermal rate constants of the four reactions in the range of temperatures between 200K and 1000K by means of a close coupling (CC) time-dependent (TDWP) approach on the new BW2 PES. [31,32]

This article is organized as follows: Section 2 gives a brief review of the theoretical methodologies to atom-diatom reactions used in the current study. The results of calculation and discussion of the result are given in Sec. 3. We conclude in

Sec. 4.

## 2. THEORY

In this section we briefly describe the time-dependent wave packet (TDWP) method employed to calculate the initial state-selected total reaction probability with the final resolved products. The reader is referred to ref. 33 for more detailed discussions of the methodology. In the present study we solve the time-dependent Schrödinger equation

$$i\hbar \frac{\partial \Psi}{\partial t} = H\Psi \qquad (1)$$

for the four Eq. X reactions. The Hamiltonian expressed in the reactant Jacobi coordinates for a given total angular momentum $J$ can be written as

$$H = -\frac{\hbar^2}{2\mu_R}\frac{\partial^2}{\partial R^2} + \frac{(\vec{J}-\vec{j})^2}{2\mu_R R^2} + \frac{\vec{j}^2}{2\mu_r r^2} + V(\vec{r},\vec{R}) + h(r), \qquad (2)$$

where $\vec{r}$ and $\vec{R}$ are the vibrational and the translational vectors, while $\mu_r$ is the reduced mass for HCl/DCl, and $\mu_R$ is the reduced mass between H/D and HCl/DCl. $\vec{J}$ and $\vec{j}$, respectively, represent the total angular momentum operator and the rotational angular momentum operator of HCl/DCl. $V(\vec{r},\vec{R})$ is the interaction potential excluding the diatomic potential of the diatom. The diatomic reference Hamiltonian $h(r)$ is defined as

$$h(r) = -\frac{\hbar^2}{2\mu_r}\frac{\partial^2}{\partial r^2} + V_r(r), \qquad (3)$$

where $V_r(r)$ is a diatomic reference potential.

The time-dependent wave function satisfying the Schrödinger equation (1) can be expanded in terms of the body-fixed (BF) translational-vibrational-rotational basis defined using the reactant Jacobi coordinates as[34]

$$\Psi^{JM\varepsilon}_{v_0 j_0 K_0}(\vec{R},\vec{r},t) = \sum_{n,v,j,K} F^{JM\varepsilon}_{nvjK,v_0 j_0 K_0}(t) u_n^v(R) \phi_v(r) Y^{JM\varepsilon}_{jK}(\hat{R},\hat{r}), \tag{4}$$

where $n$ is the translational basis label, $M$ and $K$ are the projection quantum numbers of $J$ on the space-fixed $z$ axis and body-fixed $z$ axis, respectively. $(v_0, j_0, K_0)$ denotes the initial rovibrational state, and $\varepsilon$ is the parity of the system defined as $\varepsilon = (-1)^{j+L}$ with $L$ being the orbital angular momentum quantum number. The reader can find the definitions of various basis functions elsewhere.[33]

We employed the split-operator method[34] to carry out the wave packet propagation. The time-dependent wavefunction is absorbed at the edges of the grid area to avoid artificial reflections.[35] Finally the initial state-selected total (final state-summed) reaction probabilities are obtained through the flux calculation[34] at the end of the propagation.

We construct wave packets and propagate them to calculate the reaction probabilities for each product. Using the following formula[34]

$$P^J_{v_0 j_0 K_0}(E) = \left\langle \psi^{JM\varepsilon+}_{v_0 j_0 K_0}(E) \middle| \frac{1}{2}[\delta(\hat{s}-s_0)\hat{v}_s + \hat{v}_s \delta(\hat{s}-s_0)] \middle| \psi^{JM\varepsilon+}_{v_0 j_0 K_0}(E) \right\rangle. \tag{5}$$

The initial state specific total reaction probabilities can be calculated at the end of the propagation. In the former formula, $s_0$ is the coordinate perpendicular to a surface located at flux evaluation and $v_s$ is the velocity operator corresponding to the coordinate $s$, and $\psi^{JM\varepsilon+}_{v_0 j_0 K_0}(E)$ is the time–independent wavefunction that can be obtained by Fourier transforming the TDWP wavefunction.

After the reaction probabilities $P^J_{v_0 j_0 K_0}(E)$ have been calculated for all fixed angular momenta $J$, we can evaluate the reaction cross section for a specific initial state by simply summing the reaction probabilities over all the partial waves (total angular momentum $J$). In practice, we can use the interpolation method to get the probabilities for missing values of $J$; reaction probabilities at only a limited number of total angular momentum values of $J$ need to be explicitly calculated and probabilities for missing values of $J$ are obtained through interpolation. We used the formula

$$\sigma_{v_0 j_0}(E) = \frac{\pi}{k_{v_0 j_0}^2} \sum_J (2J+1) P_{v_0 j_0}^J (E) \qquad (6)$$

where $k_{v_0 j_0} = (2\mu_R E)^{1/2}/\hbar$ is the wave number corresponding to the initial state at fixed collision energy $E$, and $P_{v_0 j_0}^J(E)$ is given by

$$P_{v_0 j_0}^J(E) = \frac{1}{2j_0+1} \sum_{K_0} P_{v_0 j_0 K_0}^J(E). \qquad (7)$$

As in refs. 13,33 we construct wave packets and propagate them to calculate the reaction probabilities for $P^J(E)$ each product. The integral cross section from a specific initial state $j_0$ is obtained by summing the reaction probabilities over all the partial waves on total angular momentum quantum number.

The calculation of the reaction rate constant for the initial states ($v=0, j=0$) of H/DCl uses a uniform version [31,32] of the $J$-shifting approach.[36] The initial state-specific thermal rate constant in the uniform $J$-shifting scheme is given

$$r'(T) = \sqrt{\frac{2\pi}{(\mu_R k_B T)^3}} Q^0(T) \sum_J (2J+1) e^{-B_J(T)J(J+1)/k_B T}. \qquad (8)$$

The shifting constant is determined by[32]

$$B_J(T) = \frac{k_B T}{J(J+1) - J_i(J_i+1)} \ln\left(\frac{Q^{J_i}}{Q^J}\right), \qquad (9)$$

where $k_B$ is the Boltzmann constant, $T$ is the temperature, and $Q^{J_i}$ is a partition-like function defined as

$$Q^{J_i} = \int P^{J_i}(E) e^{-E/k_B T} dE, \tag{10}$$

where $J_i$ is a reference angular momentum which divides total angular momentum into different ranges,[32] and $Q^J$ is similarly defined as

$$Q^J = \int P^J(E) e^{-E/k_B T} dE, \tag{11}$$

where $P^J(E)$ are the probabilities for a total angular momentum quantum number from a given initial state.[39]

The numerical parameters of the four reactions for the wave packet propagation are as follows: A total of 100 vibrational functions are employed for $r$ in the range of $[0.8, 8.5] a_0$ for the reagents HCl/DCl in the interaction region. A total number of 200 sine functions (among them 80 for the interaction region) are employed for the translational coordinate $R$ in a range of $[0.8, 14.0] a_0$.[39] For the rotational basis we use $j_{max}=45$. The number of $K$ used in our calculation is given by $K_{max}$ = max (3, $K_0+2$) starting with $K_0=0$. The largest number of $K$ used is equal to 6 for the $j=0$, $K_0=0$ initial state (for $\varepsilon=-1$, there is one less $K$ block used). These values of $K_0$ and $K_{max}$ were determined following an extensive series of tests.[1] The initial wave packet was centered at R=10 $a_0$, with a width of 0.23 $a_0$ and average translational energy of 0.8 eV.[39] It was found that convergence of total cross sections, for all the reported initial (rotational) states in the entire energy region, was achieved up to a few percent. For lower $J$, we propagate the wave packets for 15000a.u. of time to converge the low energy reaction probability (in all calculations a time step-size of 10a.u. was used). In this calculation, we used $J$ from 0 to 80 to calculate the cross section. For $J>20$, we propagate the wave packets for a shorter time because the reaction probability in the low energy region is negligible.[34]

## 3. RESULTS AND DISCUSSION

### A. Reaction probabilities

First of all, we computed the energy resolved reaction probability for the Eq. X reactions for collision energies at the range of [0.1,1.4]eV with HCl/DCl initially in their ground states. We plotted the results of probabilities of $J=0$ as a function of the collision energy on BW2 potential for all possible reactions in Fig.1. As shown in Fig.1, the behavior of the reaction probabilities corresponding to Eq. X reactions are quite different. For the exchanged channel in Fig.1(a), the difference seems to increase with the increasing of the collision energy. The reaction probabilities for the H+HCl reaction are systematically lower than those of H+DCl reaction. Moreover, for the abstraction channel one can find from Fig.1(b) that the reaction probabilities for the D+DCl reaction are very close to those of the H+HCl reaction. The reaction probabilities of the H+DCl are systematically lower than those of the other three reactions. The reaction probability of the D+HCl reaction is much higher than the other three reactions in both channels. Perhaps the high zero-point energy made the D+HCl reaction probabilities higher than the others.

The H/D+HCl reaction exhibits a strong oscillating effect, but it is smooth for H/D+DCl. That is because the tunneling effect affects the H atom stronger than the D atom. As a result, the oscillating behavior displayed in the four reactions nearly disappears for the abstraction channel.

The reaction probabilities as a function of collision energy for total angular momentum of $J=0,8,18$ and 28, are presented graphically in Fig.2 and 3 of the four reactions for both channels. It can be seen that in Fig. 2 the curve of the reaction probabilities of D+DCl are very similar to the D+HCl reaction. The probabilities of $J=18,28$ are higher than the $J=0,8$ in the high energy. And the H+DCl and H+HCl are very similar. The probabilities of $J=28$ are almost equal to zero in the bottom two figures of Fig. 2. It can be seen from the Fig. 3 abstraction channel that no significant oscillating behavior is found. The values of reaction probabilities decrease with an increase of $J$ in the abstraction channel. The threshold of probabilities increases with the increasing of $J$ in both channels.

The effect of the initial reagent rotation excitation on the reaction probability for the four reactions ($J=0, v=0, j=1,3,5,7,9$) is shown in Figs. 4,5,6,7. As seen from the figures, the reaction probabilities of the four reactions increased obviously and their oscillating becomes stronger as the rotational quantum number $j$ increases in the exchanged channels.[37] The increase and the oscillating perhaps can be explained by a long-range van der Waals well in both the entrance and exit channel that is the same as the explanation in ref. 18. The negative values of the reaction probabilities in the low energy region are negligible with the approximation of the theory.

**B. Integral Cross Sections**

Next we calculate the integral cross section for the initial ground state of HCl/DCl on a BW2 surface. The calculated integral cross sections for the four reactions are depicted in Fig.8 for the both channels. Fig. 8(a) is for the exchanged channel, and Fig. 8(b) is for the abstraction channel. As can be seen from Fig. 8(a), the cross sections of the four reactions increase with further increase of the collision energy. It can be seen that the four reactions curves are very similar, but the cross section of the H+HCl reaction are the highest of the four reactions for the exchanged channel, while those of D+HCl reaction is the highest one for the abstraction channel.

The calculated cross sections for H+DCl are depicted in Fig. 9. As can be seen, the present cross sections for the abstraction channel are systematically larger than the corresponding experimental values (triangle) of Polanyi and co-workers,[26] and the results of QCT are bigger than the result of TDWP except for the range of collision energy [0.8,1.2]eV, and they have the same threshold energy. But for the exchanged channel the cross sections are almost equal to the corresponding experimental value (squares) lying between 0.25Å$^2$ and 0.75 Å$^2$ for the collision energy range considered in previous calculations.[26] The result of QCT is much lower than that of TDWP in the lower energy but after 1.35eV it changes. The threshold energy of QCT is much higher than that of TDWP, because the TDWP method has included the zero-point energy and the tunneling effect.

The four reactions have a slightly different threshold energy for the same channels. The difference can be interpreted by the different zero-point energy and the different tunneling effect. In the collinear transition state for the abstraction reaction, the height of the classical barriers is 0.184eV for BW2 PES.[29] For the H+ClH exchange reaction, which also has a collinear transition state, the barrier height is computed to be 0.776eV for BW2 PES.[29] The exchanged channel of the BW2 surface also exhibit small vdw wells. The barrier heights of the abstraction channel are much lower than the exchanged channel of the four reactions. So the thresholds of the four reactions for an exchanged channel are a little bit higher than those of the abstraction channel. Overall the results calculated for the abstraction channel are much higher than those given for the exchange channel on a BW2 surface.

## C. Rate constant

To calculate the total reaction probability for more than two values of $J$ in order to obtain a more accurate rate constant,[32] a very accurate rate constant can be obtained by using reaction probabilities evaluated at more than 6 values of $J$.

We calculate the reaction rate constants $r'(T)$ that are given for the initial states by using the uniform version of the $J$-shift approach. The calculation of the thermal rate constants is in the range of temperatures between 200K and 1000K. The calculated rate constants for the four reactions are depicted in Fig. 10 for both channels. Fig. 10(a) is for the exchanged channel and Fig. 10(b) is for the abstraction channel. As can be seen, from the Fig. 10(a) at lower temperature the present rate constants of D+DCl are the highest, and those of the H+HCl are the smallest. While at the higher temperature the present rate constant of H+DCl is the highest one, and that of the D+HCl is the smallest one. From Fig. 10(b) one can see that the rate constant curves of the H+HCl reaction are almost equal to those of D+HCl reaction. And the rate constants of H+DCl and D+DCl are almost equal. The rate constants of the H/D+DCl reactions are smaller than those of H/D+HCl reactions. As can be seen, the present rate constants roughly follow the same as the corresponding experimental values which are in the range of $[10^{-17},10^{-12}]$ cm$^3$s$^{-1}$.[30-37]

This can also be taken as an indication that the existence of quantum effects such as tunneling may play an important role. So, according to our calculation results here, the possible reasons for the nonlinear behavior of the Arrhenius plots of the reactions should be the combination influence of tunneling effect.

## 4. CONCLUSIONS

In this work we have investigated the reaction probabilities as a function of the collision energy for the four H+HCl, H+DCl, D+HCl and D+DCl reactions and studied the influence of the initial state excitation of the reagents applying the TDWP method to study the H+HCl reaction on the BW2 PES. First, the quantum tunneling can strengthen reactive ability of H abstraction in low temperatures and translational energy, and the deuterium replacement of hydrogen will weaken the tunneling effect. In the high temperature and collision energy condition, the H atom has more chances to collide with a chlorine atom and thus the reaction will produce more HCl, so the exchanged channel increases faster than the abstraction channel. Second, the van der Waals interactions and the centrifugal barrier, which traps the hydrogen for a finite time, are included for depicting the effect of the rotational quantum number $j$ for the four reactions. Third, the cross sections and rate constants are calculated and explained by the zero-point energy and the tunneling effect for the four reactions, which are higher than the previous results. The agreement of the presented results is good enough comparing with the experimental results, but the BW2 PES need to be improved.

**Acknowledgment**

This work is supported by NSFC (Grants Nos. 29825107 and 29853001) and NKBRSF as well as the Knowledge Innovation Program of the Chinese Academy of Sciences (Grant: DICP K2001/E3).

**Figure Captions**

Fig.1. Total reaction probabilities for $J=0$ from the ground state of the HCl/DCl reactant for the four reactions Eq. X on the BW2 PES. Figure (a) is for exchanged channel. Figure (b) is for the abstraction channel. The solid line is for D+DCl, dashed line is for D+HCl, dotted line is for H+DCl, dashed-dotted line is for H+HCl.

Fig.2. Total reaction probabilities for $v=0, J=0,8,18,28$ of the HCl/DCl reactant for exchange channel on the BW2 potential. The solid line is for all the $J=0$, dashed line is for $J=8$, dotted line is for $J=18$, dashed-dotted line is for $J=28$.

Fig.3. Same as Fig.2 but for the abstraction channel.

Fig.4. Total reaction probabilities for $J=0$, $v=0$ from the ground state of the DCl reactant for D+DCl→D+DCl (a) D+DCl→$D_2$+Cl (b) on the BW2 potential. The solid line is for $j=1$, dashed line is for $j=3$, dotted line is for $j=5$, dashed-dotted line is for $j=7$, dashed-dotted-dotted line is for $j=9$.

Fig.5. Same as Fig.4 but for the D+HCl→H+DCl(a) D+HCl→DH+Cl (b) reaction.

Fig.6. Same as Fig.4 but for the H+DCl→D+HCl (a) H+DCl→HD+Cl (b) reaction.

Fig.7. Same as Fig.4 but for the H+HCl→H+HCl(a) H+HCl→$H_2$+Cl (b) reaction.

Fig.8. Reaction Cross section for $v=0$, $j=0$ from $J=0$ to 80 of the HCl/DCl reactant for the four reactions Eq. X on the BW2 potential. The left figure is for the exchanged channel. The right figure is for the abstraction channel. The solid line is for D+DCl, dashed line is for D+HCl, dotted line is for H+DCl, dashed-dotted line is for H+HCl.

Fig.9. Reaction cross section as a function of collision energy for H+DCl reaction using TDWP approach and QCT on the BW2 PES. The triangle is the experimental results of ref. 26, 40 for the abstraction channel of H+DCl reaction while the squares correspond to the experimental result of the exchanged channel of ref. 26, 40 for the reaction. The solid line and the dashed line are calculated for the TDWP approach. The dotted line and the dashed-dotted line are calculated for QCT.

Fig.10. The Rate Constant for $v=0, j=0$ of the HCl/DCl reactant for the four reactions Eq. X on the BW2 potential. The left figure is for the exchanged channel. The right figure is for the abstraction channel. The solid line is for D+DCl, dashed

line is for D+HCl,dotted line is for H+DCl, dashed-dotted line is for H+HCl.

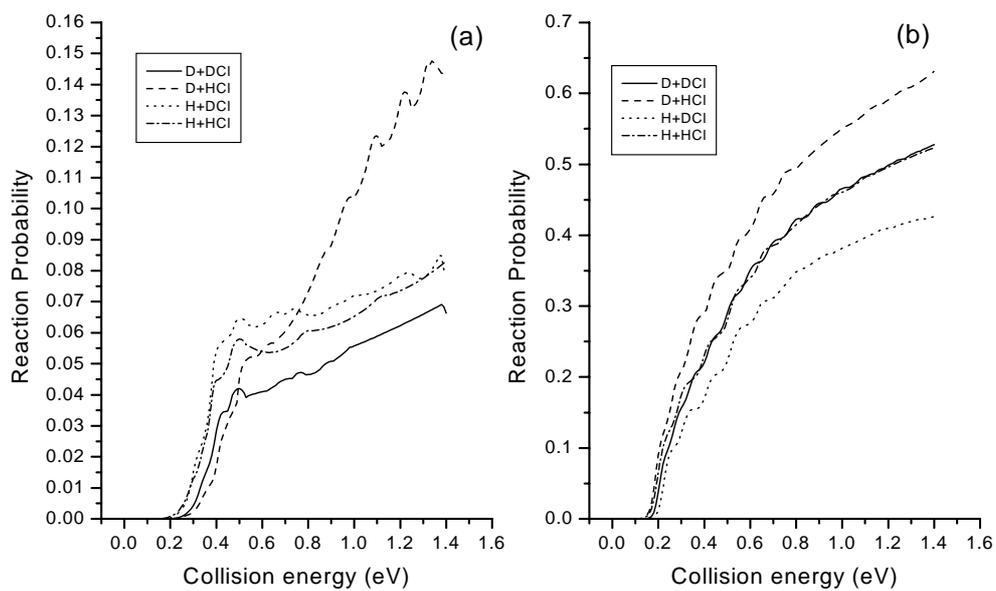

Figure 1

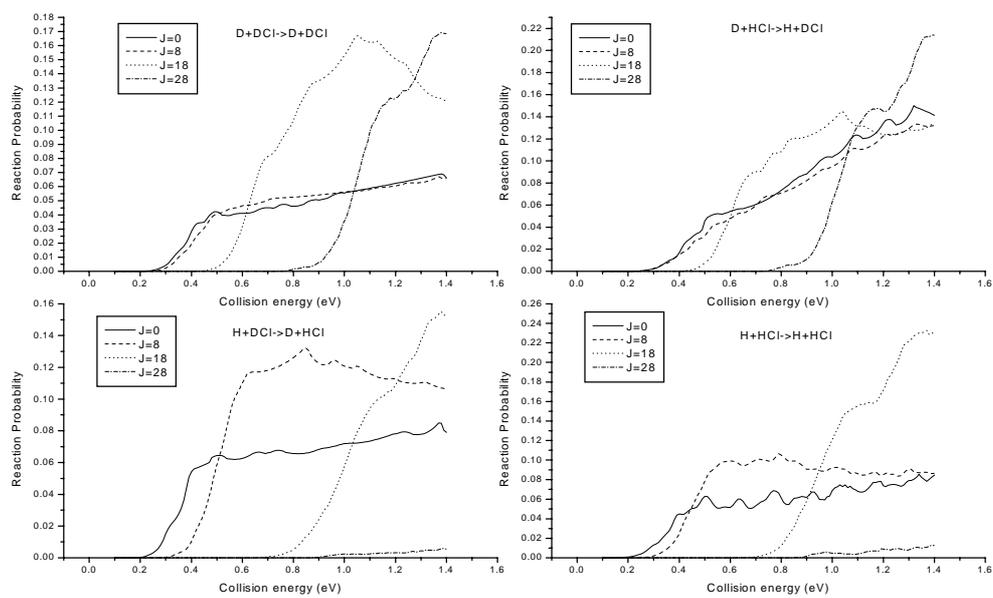

Figure 2

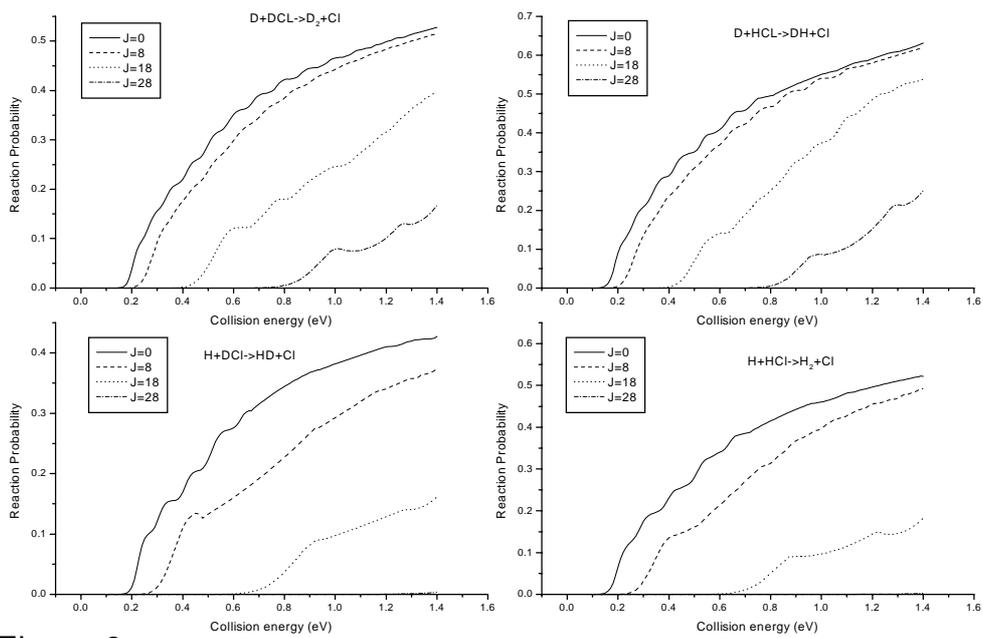

Figure 3

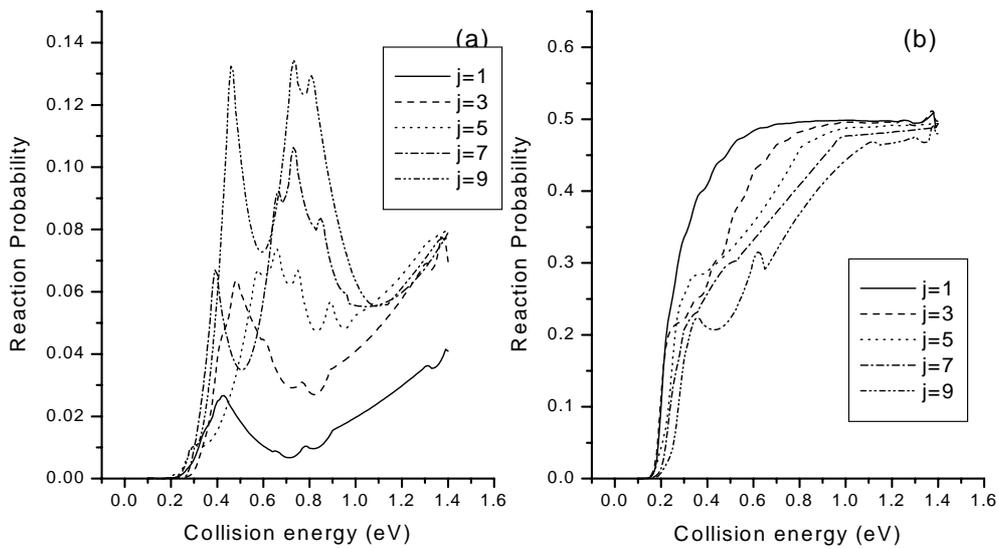

Figure 4

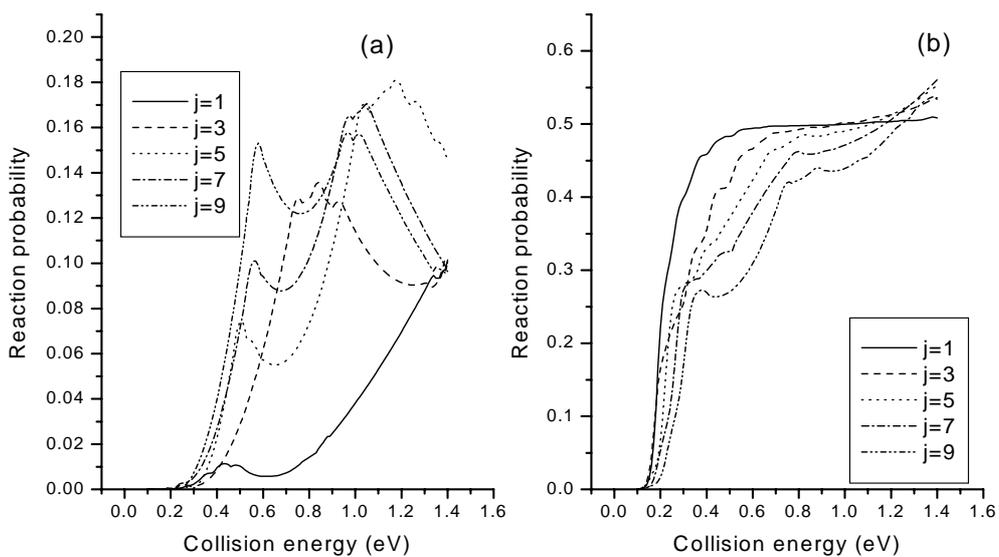

Figure 5

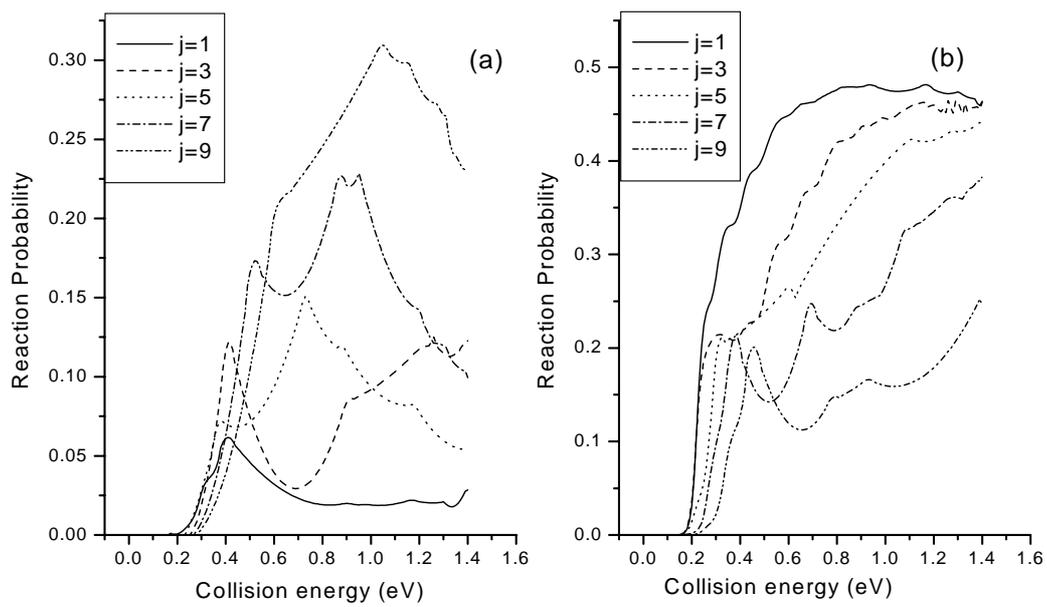

Figure 6

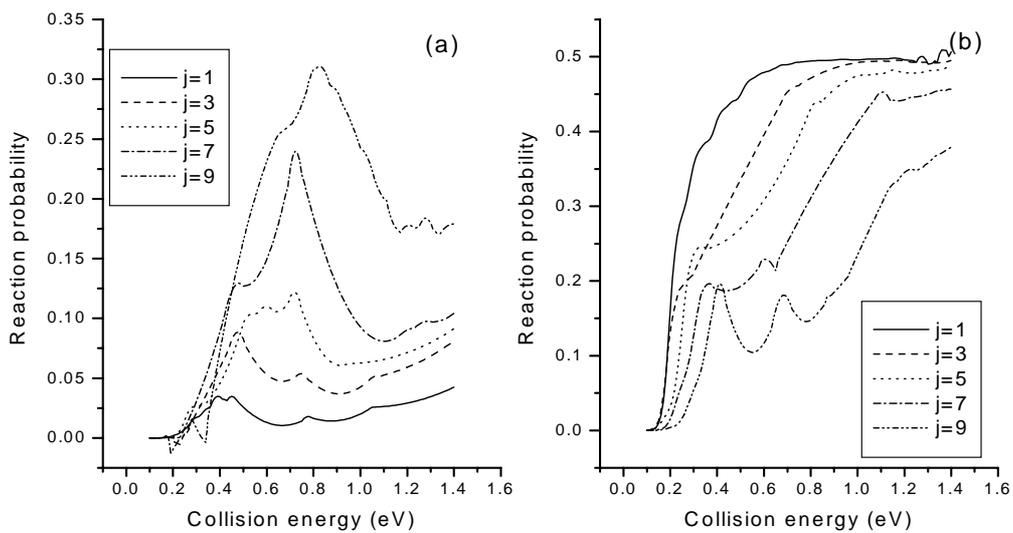

Figure 7

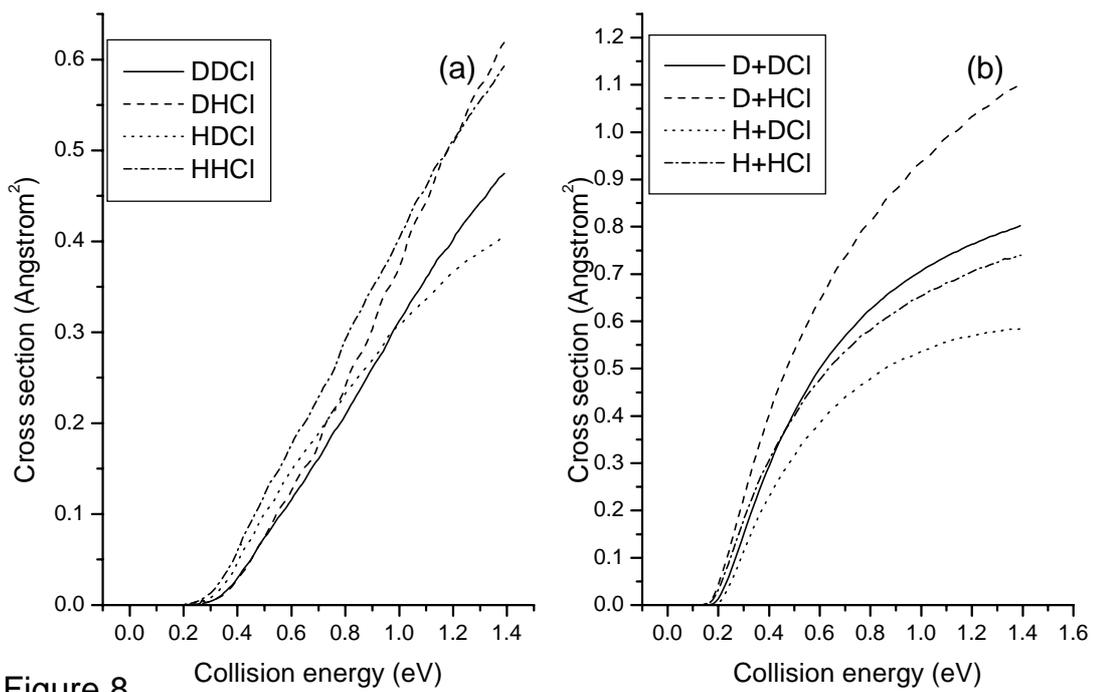

Figure 8

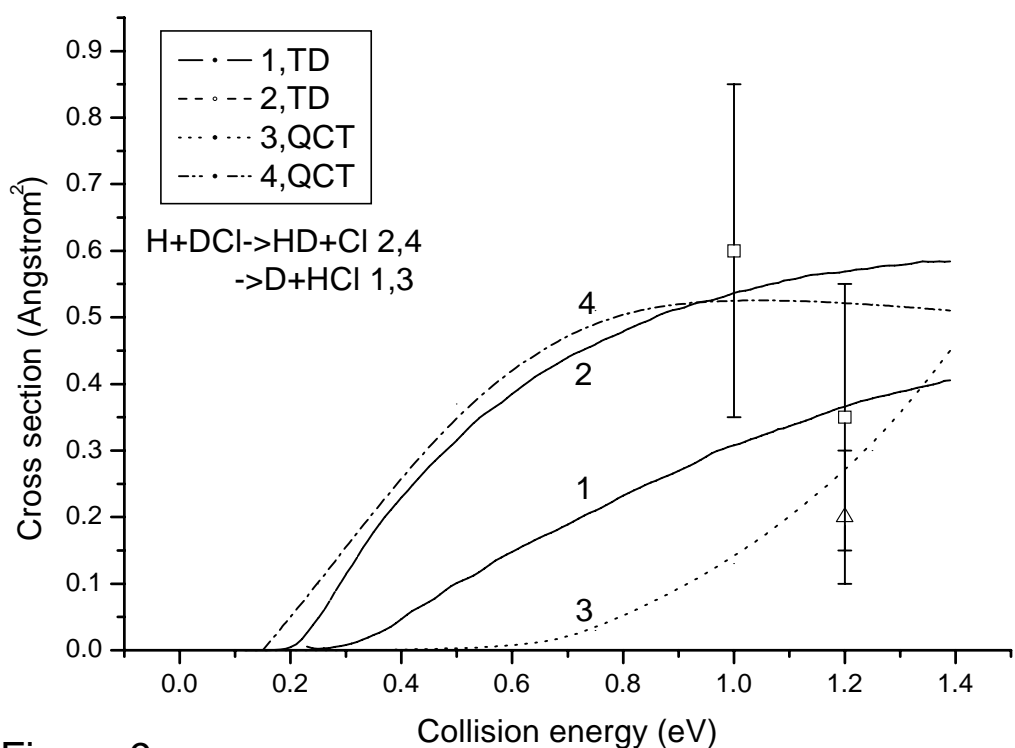

Figure 9

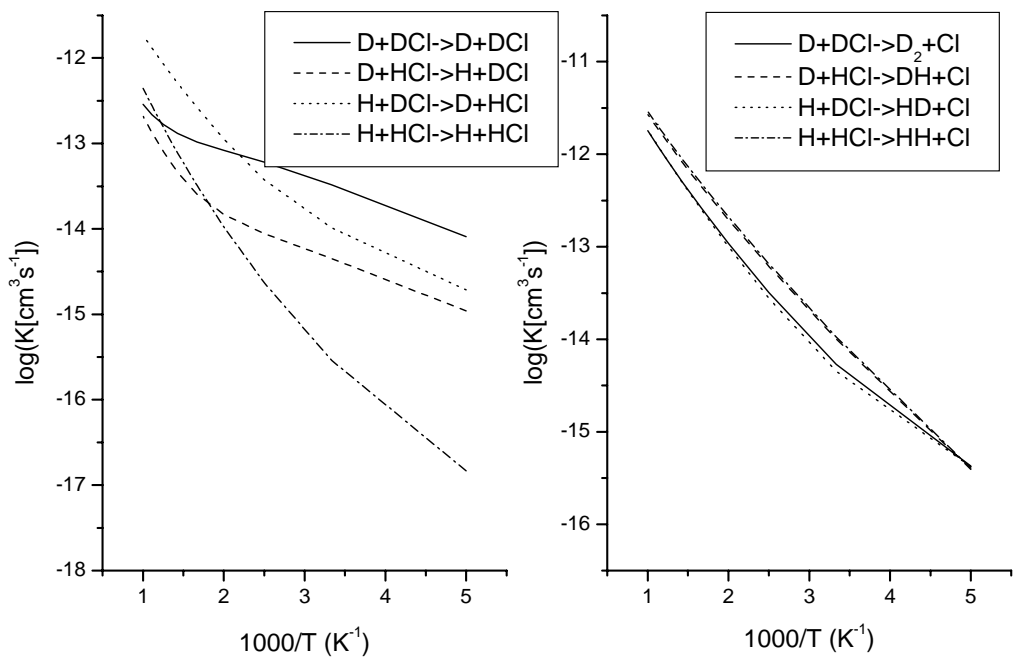

Figure 10